# Electron acoustic solitary wave in quantum plasmas with Kappa electrons

Aakanksha Singh and Punit Kumar

*Deapertment of Physics, University of Lucknow, Lucknow-226007, India*

kumar_punit@lkouniv.ac.in

## ABSTRACT

Electron-acoustic solitary waves (EASWs) in quantum plasma comprising stationary ions, cold electrons, hot electrons, and kappa-distributed electrons have been investigated. The generalized Kappa-Fermi distribution has been modified to include electrostatic energy contribution and the number density of Kappa electrons has been obtained using this modified distribution. Utilizing the quantum hydrodynamic (QHD) model, a dispersion relation has been derived for linear EAWs. Employing the standard reductive perturbation technique, a Korteweg–de Vries (KdV) equation governing the dynamics of EAWs have been derived. The quantum mechanical effects of different parameters like kappa index, Mach number and equilibrium kappa electron density have been examined on the profiles of EASWs. It is found that the presence of kappa electrons in quantum plasma leads to new results, including steeper dispersion curves, sharper and more localized solitary waves with kappa index and stronger plasma interactions with increased kappa electrons density in dense astrophysical environment.

**Keywords:** Quantum plasma, Electron acoustic waves (EAWs), Kappa-distribution, Solitons, Korteweg-de Vries (KdV), Dispersion, QHD model, Nonlinear dynamics.

# 1. Introduction

Quantum plasmas have attracted significant attention due to their unique properties and their relevance to various physical systems, including quantum wells [1], semiconductors [2], thin films, nanometallic structures [3], ultracold plasmas [4], intense laser-solid interaction experiments [5], x-ray scattering spectral measurements [6], and dense astrophysical objects such as white dwarfs and neutron stars [7–9]. Unlike classical plasmas, which are characterized by low densities and high temperatures, quantum plasmas are defined by high densities and low temperatures. In white dwarf interiors and neutron star crusts, the plasma is extremely dense and highly degenerate [10,11]. At such high particle densities, the Fermi temperature typically exceeds the system's thermal temperature, and the de Broglie wavelength of charge carriers becomes comparable to their interparticle spacing [12]. Under these conditions, the plasma exhibits behaviour similar to a Fermi gas, with quantum mechanical effects significantly influencing the dynamics of charged particles [13–15].

Electron acoustic waves (EAWs) are high-frequency dispersive plasma waves (relative to the ion plasma frequency) in which a small population of inertial cold electrons oscillates against a dominant population of inertialess hot electrons [16–18]. EAWs have been observed in laser produced plasma experiments and dense astrophysical plasmas [19, 20], where the presence of two distinct electron populations is evident. In last few years, the linear as well as nonlinear propagation of EAWs in unmagnetized and magnetized quantum plasmas with planar and nonplanar geometries has been studied extensively [21–24]. Several studies have been performed on EASWs in dense astrophysical plasma using QHD model [25-31].

The theoretical studies on EAWs, in quantum plasmas are mainly based on models that deal with the standard Fermi distribution function for electrons. However, in quantum astrophysical plasmas, which are characterized by extremely high particle densities and Fermi temperatures significantly exceeding the thermal temperature of the environment $\beta \varepsilon_F \gg 1$, particle distributions may deviate from the standard Bose-Einstein or Fermi-Dirac distributions. In such cases, specific populations of particles can be described using quantum versions of the Kappa distribution: the Kappa-Bose and Kappa-Fermi distributions, also referred to as the Olbert-Bose and Olbert-Fermi distributions, respectively [32]. In this paper, we focus on the Kappa-Fermi distribution as this one is more important in view of the wide range of quantum applications. The Kappa Fermi distribution consists of the power law tail and the related stretching of the gap [32]. In the case of finite temperature $\beta \ll \infty$, an energy gap appears which to higher energy turns into a power law decay of the distribution, where power is a function of Olbert parameter κ and the thermodynamic constant (s).

All theoretical investigations of Electrostatic Acoustic Waves (EAWs) have so far been conducted using the classical Kappa distribution function [33-38], primarily within the context of space and astrophysical plasmas [39-49]. However, no research has yet examined EAWs in quantum plasmas utilising the quantum Kappa distribution. Therefore, the application of the quantum Kappa distribution in the analysis of EAW dynamics, particularly in dense astrophysical plasmas, represents a novel aspect of the present research.

In the present work, we study the linear and nonlinear dynamics of EAWs in collisionless quantum plasma consisting of two different temperature electron fluid (cold and hot) and immobile ion in the presence of Kappa Fermi distributed hot electrons. The generalized Kappa-Fermi distribution has been modified to include electrostatic energy contribution and the number density of Kappa electrons has been obtained using this modified distribution. The theory of this paper has been built using the quantum hydrodynamic (QHD) model. This model can effectively examine the behaviour of quantum plasma constituents and supports the study of nonlinear phenomena and collective excitations

[50] in quantum plasmas. The QHD model has proven its productiveness within astrophysical objects, where quantum effects are crucial for describing degenerate electrons and high-density ion dynamics, such as white dwarfs, neutron stars [51,52] and in stellar evolution [53]. The Korteweg-de Vries (KdV) equation has been employed to describe the Solitary wave propagation in unmagnetized plasmas without the dissipation and geometry distortion. The KdV equation provides a simplified framework for analyzing the dynamics of EASWs, providing a deeper understanding of their formation, propagation, and dependence on plasma conditions. Further, the effects of equilibrium Kappa electron density ($n_{\kappa e0}$), Kappa index ($\kappa$), quantum diffraction parameter ($H$) have been studied on dispersion properties as well as the effect of these parameter with the speed of travelling wave ($U$) have been also studied on the solution of EASWs.

The remaining part of this paper is organized as follows: In Sec. II, we have derived the generalized form of Kappa distribution function and electron number density. In Sec. III, we considered model equations. A linear dispersion relation is derived in Sec. IV. In Sec. V we presented the nonlinear analysis and soliton solution. Finally, in Sec. VI we present summary along with discussion.

## 2. The Kappa - Fermi distribution

We consider the four component dense quantum plasma consisting of three population of electrons, i.e., inertial cold electrons, inertialess hot electrons, Kappa distributed electrons and stationary ions forming the neutralizing charge background. These fours species are denoted by c, h, $\kappa$ and i, respectively. The Kappa-Fermi distribution for degenerate Olbert Fermi gases with momentum $p$ and particle energy $\varepsilon_p = p^2/2m$ is given as [32],

$$f_\kappa(\varepsilon_p) = A\left[1 + \left\{1 + \frac{\beta}{\kappa}(\varepsilon_p - \mu)\right\}^{\kappa+s}\right]^{-\left(1+\frac{1}{\kappa+s}\right)}, \quad \varepsilon_p > \mu \tag{1}$$

where, A is a normalization constant, $T \equiv \beta^{-1}$ is a constant physical temperature (in energy units) and $\mu > 0$ represents the chemical potential with positive values, applicable to the Fermi-distribution. The Olbert parameter $\kappa$ accounts for deviations from the standard Fermi distribution, reflecting the influence of internal correlations, or additional degrees of freedom, where applicable. For $\kappa \to \infty$, the Kappa-Fermi distribution function given in eq. (1) reduces to the standard Fermi-Dirac distribution [32]. This distribution retains the quantum properties of the gas, which are particularly significant at low temperatures. Here, s > 1 is a fixed constant number that is introduced to account for thermodynamic considerations. The specific values of this constant number are determined by the nature of the gas: for ideal non-relativistic gases s = 5/2, and for ideal relativistic gases s = 4, in accordance with the thermodynamic constraints [54,55].

Taking normalization of $f_\kappa(\varepsilon_p)$ over momentum space such that $\int f_\kappa(\varepsilon_p) d^3p = n_{jo}$, we obtain the following Kappa-Fermi distribution function,

$$f_\kappa(\varepsilon_p) = \frac{n_{jo}\left\{2 - \mu\beta\left(1+\frac{s}{\kappa}\right)\right\}^{\left(\frac{1}{\kappa+s}-\frac{1}{2}\right)} \Gamma\left(\frac{1}{\kappa+s}+1\right)}{\pi^{3/2}\left[\frac{\left(1+\frac{s}{\kappa}\right)\beta}{2m_j}\right]^{-3/2} \Gamma\left(\frac{1}{\kappa+s}-\frac{1}{2}\right)} \left[1 + \left\{1 + \frac{\beta}{\kappa}\left(\frac{p^2}{2m_j} - \mu\right)\right\}^{\kappa+s}\right]^{-\left(1+\frac{1}{\kappa+s}\right)},$$

(2)

where, $n_{jo}$ is the equilibrium number density, $m_j$ is the mass of the fermion-species $j$ (for example, $j = e^-, e^+$ etc). In case of presence of electrostatic potential ($\phi$), we have to modify the above distribution to include electrostatic energy contribution. Therefore, using the energy conservation relation $\frac{p^2}{2m_j} = \frac{p_j^2}{2m_j} + q_j\phi$, where $q_j\phi$ is the increase in potential energy due to presence of electrostatic potential, $q_j$ is the charge of species $j$ and $p$ is the momentum of the particles in the initial equilibrium state. Hence, generalized form of three dimensional Kappa- Fermi distribution in the presence of electrostatic potential is

$$f_\kappa(\phi) = \frac{n_{jo}\left\{2-\mu\beta\left(1+\frac{s}{\kappa}\right)\right\}^{\left(\frac{1}{\kappa+s}-\frac{1}{2}\right)} \Gamma\left(\frac{1}{\kappa+s}+1\right)}{\pi^{3/2}\left[\frac{\left(1+\frac{s}{\kappa}\right)\beta}{2m_j}\right]^{-3/2} \Gamma\left(\frac{1}{\kappa+s}-\frac{1}{2}\right)} \left[1+\left\{1+\frac{\beta}{\kappa}\left(\frac{p_j^2}{2m_j}+q_j\phi-\mu\right)\right\}^{\kappa+s}\right]^{-\left(1+\frac{1}{\kappa+s}\right)}.$$

(3)

Integrating the above Kappa-distribution over momentum space, we can obtain the number density for the degenerate Olbert-Fermi gas showing the dependency on electrostatic potential as,

$$n_j(\phi) = \frac{n_{jo}}{4\pi^3 \hbar^3 A_\kappa}\left[\frac{1+(q_j\phi-\mu)\left(1+\frac{s}{\kappa}\right)\beta}{1-\mu\beta\left(1+\frac{s}{\kappa}\right)}\right]^{\frac{1}{2}-\frac{1}{\kappa+s}}.$$

(4)

Thus, the number density of kappa electrons ($q_j = -e$) can be written as

$$n_{\kappa e}(\phi) = \frac{n_{\kappa e 0}}{4\pi^3 \hbar^3 A_\kappa}\left[\frac{1-(e\phi+\mu)\left(1+\frac{s}{\kappa}\right)\beta}{1-\mu\beta\left(1+\frac{s}{\kappa}\right)}\right]^{\frac{1}{2}-\frac{1}{\kappa+s}},$$

(5)

where, $\mu = \frac{A_\kappa^{2/3}\hbar^2 k_F^2}{2m}$ represents the chemical potential for Fermi gas, $n_{\kappa e 0}$ is the equilibrium density of Kappa distributed electrons. $A_\kappa = 2^{1/(\kappa+s)}$ is modification term which depends upon

Kappa parameter in case of Olbert-Fermi gas with Olbert Fermi momentum $P_{OF} = A_\kappa^{1/3} p_F$, $p_F = \hbar k_F$ is the Fermi momentum and $k_F = \left(3\pi^2 n_{\kappa e0}\right)^{1/3}$ is the Fermi wavenumber. For large κ ($\kappa \gg 1$), there is no significant deviation i.e., the distribution becomes indistinguishable from the ordinary Fermi-Dirac form. At moderate κ > 1, the effective temperature in the distribution is effectively raised, which extends the domain of finite-temperature effects, even at low temperatures. In contrast, for small values of κ (κ < 1), this effect is reversed with fractional values of κ lowering the effective temperature. Therefore, the impact of the Olbertian transformation on the properties of a degenerate Fermi gas is relatively mild and becomes significant primarily for small κ (κ < 1), where it is mainly determined by the value of $s = 5/2$ and remains almost constant. As κ approaches zero and $s = 5/2$, the normalization constant ($A_\kappa$) tends towards approximately 1.36. In this sense, the role of Kappa parameter (κ) is, to either stretch or compress the quantum domain of the ideal gas [32].

## 3. Governing set of equations

In EAWs, the cold electrons provide the inertia and hot electrons the restoring force, respectively. The phase speed of the EAW lies in the range $v_{Fce} \ll \omega/k \ll v_{Fhe}$, where $v_{Fce}$ and $v_{Fhe}$ are the Fermi velocities of cold and hot electrons, respectively. EAW propagates on cold electron dynamic scale and $n_{ce0} \ll n_{he0}$, therefore $\omega_{pce} \ll \omega_{phe}$ holds. The plasma frequency due to hot and cold electrons is defined as $\omega_{pe\alpha} = \left(n_{\alpha e0} e^2 / m_e\right)^{1/2}$. The electron acoustic speed is defined as $c_{ea} = \left(2 k_B T_{Fhe} / m_e \delta\right)^{1/2}$ where, $\delta = n_{he0}/n_{ce0} > 1$ and $\lambda_{Fhe} = \left(2 K_B T_{Fhe} / n_{he0} e^2\right)^{1/2} = v_{Fhe}/\omega_{phe}$ is the Fermi wavelength due to hot electrons in quantum plasma. The basic set of equations describing the dynamics of EAWs in such plasma given are [21,22],

$$\frac{\partial n_\alpha}{\partial t} + \frac{\partial}{\partial x}(n_\alpha \vec{v}_\alpha) = 0, \tag{6}$$

$$\frac{\partial \vec{v}_\alpha}{\partial t} + \vec{v}_\alpha \frac{\partial \vec{v}_\alpha}{\partial x} = \frac{e_\alpha}{m_\alpha}\frac{\partial \phi}{\partial x} - \frac{1}{m_\alpha n_\alpha}\frac{\partial P_{F\alpha}}{\partial x} + \frac{\hbar^2}{2m_\alpha^2}\frac{\partial}{\partial x}\left(\frac{1}{\sqrt{n_\alpha}}\frac{\partial^2}{\partial x^2}\sqrt{n_\alpha}\right), \tag{7}$$

and

$$\frac{\partial^2 \phi}{\partial x^2} = \frac{e}{\varepsilon_0}(n_c + n_{he} + n_{ke} - Z_i n_{0i}). \tag{8}$$

Eqs. (6) and (7) correspond to continuity and momentum equations for $\alpha$- species ($\alpha = c, h$) respectively. The second term on left hand side of eq. (7) is the convective derivative of the velocity. The first term on RHS of eq. (7) is the force under the influence of electrostatic field as $E = -\nabla\phi$, where $\phi$ is the electrostatic potential. The second term is the force due to Fermi pressure $P_{F\alpha} = m_\alpha v_{F\alpha}^2 n_\alpha^3 / 3n_{0\alpha}^2$, where $v_{F\alpha} = \sqrt{2k_B T_F / m_{e\alpha}}$ is Fermi velocity and $k_B$ is the Boltzmann constant, the third term corresponds to the force of quantum Bohm potential arising from quantum corrections in the density fluctuations. $\hbar$ is reduced Planck's constant. The densities of the Kappa distributed hot electrons, inertial cold electrons, non inertial hot electrons and the stationary ions are coupled by Poission equation (8). In equilibrium, the plasma holds quasi-neutrality condition, $n_{ce0} + n_{he0} + n_{\kappa e0} = z_i n_{i0}$ where $n_{ce0}$, $n_{he0}$ and $n_{\kappa e0}$ are equilibrium densities of hot, cold and kappa electrons. Here $n_{i0}$ is the background ion density.

The Fermi temperature is directly related to electron density in dense plasmas. Since the condition $n_{ce0} \ll n_{he0}$ must hold for the electrostatic wave (EAW), it follows that $T_{Fce} \ll T_{Fhe}$ in quantum plasmas, the Fermi pressure contributed by cold electrons can be neglected relative to the pressure from hot electrons in the model. Additionally, since the phase speed of the EAW lies within a certain range $v_{Fce} \ll \omega/k \ll v_{Fhe}$, the inertia of the hot electrons can be considered negligible in this model. Thus, from eq. (7), momentum equation for cold electrons and inertialess hot electron can be written as,

$$\frac{\partial \vec{v}_c}{\partial t} + \vec{v}_c \frac{\partial \vec{v}_c}{\partial x} = \frac{e}{m_e}\frac{\partial \phi}{\partial x} + \frac{\hbar^2}{2m_e^2}\frac{\partial}{\partial x}\left(\frac{1}{\sqrt{n_{ce}}}\frac{\partial^2}{\partial x^2}\sqrt{n_{ce}}\right), \tag{9}$$

and

$$0 = \frac{e}{m_e}\frac{\partial \phi}{\partial x} - \frac{1}{m_e n_{he}}\frac{\partial P_{Fhe}}{\partial x} + \frac{\hbar^2}{2m_e^2}\frac{\partial}{\partial x}\left(\frac{1}{\sqrt{n_{he}}}\frac{\partial^2}{\partial x^2}\sqrt{n_{he}}\right). \tag{10}$$

For hot electrons, the equation of state is described by the one dimensional Fermi-gas model which is given as, $P_{Fhe} = m_{he} v_{Fhe}^2 n_{he}^3 / 3 n_{he0}^2$ [56]. This equation of state of hot electrons is the same as in ordinary metals for which the electron Fermi temperature is generally much higher than the room temperature. Therefore, the pressure term in eq. (10) equation can be expressed as, $\frac{1}{m_{he} n_{he}}\frac{\partial P_{Fhe}}{\partial x} = \frac{v_{Fhe}^2}{n_{he0}^2} n_{he} \frac{\partial n_{he}}{\partial x}$. Additionally, the non-dimensional quantum parameter is, $H = \hbar \omega_{pe\alpha} / m_{e\alpha} v_{Fe\alpha}^2$. So, the general Bohm potential term in eq. (9) and (10) can be written as, $\frac{\hbar^2}{2m_\alpha^2}\frac{\partial}{\partial x}\left(\frac{1}{\sqrt{n_\alpha}}\frac{\partial^2}{\partial x^2}\sqrt{n_\alpha}\right) = \frac{H_\alpha^2 v_{F\alpha}^4}{4\omega_{P\alpha}^2 n_{0\alpha}}\left(\frac{\partial^3 n_\alpha}{\partial x^3}\right)$. By using these representations, eqs (9) and (10) become,

$$\frac{\partial \vec{v}_c}{\partial t} + \vec{v}_c \frac{\partial \vec{v}_c}{\partial x} = \frac{e}{m_e}\frac{\partial \phi}{\partial x} + \frac{H_{ce}^2 v_{Fce}^4}{4\omega_{Pce}^2 n_{ce0}}\left(\frac{\partial^3 n_c}{\partial x^3}\right), \tag{11}$$

and

$$0 = \frac{e}{m_e}\frac{\partial \phi}{\partial x} - \frac{v_{Fhe}^2}{n_{he0}^2} n_{he} \frac{\partial n_{he}}{\partial x} + \frac{H_{he}^2 v_{Fhe}^4}{4\omega_{Phe}^2 n_{he0}}\left(\frac{\partial^3 n_{he}}{\partial x^3}\right). \tag{12}$$

We assume the presence of a sufficient population of hot electrons characterized by a Kappa parameter $\kappa < 1$ and described by the Kappa-Fermi distribution or $\kappa$ - Fermi distribution, as shown in equation (4). The κ-parameter reflects the deviation of these electrons from the standard Fermi distribution. The electron density for this distribution is given by the expression in equation (5). Hence, the Poission equation (8) becomes

$$\frac{\partial^2 \phi}{\partial x^2} = \frac{e}{\varepsilon_0}\left[ n_{ce} + n_{he} + \alpha_\kappa \left\{ 1 - \left(\frac{\kappa + s - 2}{4\kappa}\right)\beta \right\} - Z_i n_{i0} \right]. \tag{13}$$

where, $\alpha_\kappa = \frac{n_{\kappa e 0}}{4\pi^3 \hbar^3 A_\kappa}\left[1 - \frac{\mu\beta}{2}\left(1 + \frac{s}{\kappa}\right)\right]^{\frac{1}{\kappa+s}-\frac{1}{2}}$.

Equations (6), (11), (12) and (13), referred as the four set of governing equations describing the dynamics of the plasma system. Eq. (6) corresponds to continuity equation for cold as well as hot electrons. Eq. (11) is the momentum equation for inertial cold electron. The second term on left hand side of eq. (11) is the convective derivative of the velocity, the first term on RHS represents the force under the influence of electrostatic field and the second term on RHS corresponds to the force of quantum Bohm potential. Eq. (12) is the equation of motion for inertialess hot electron in which the second term on the RHS is the force due to Fermi Pressure. Equation (13) corresponds to Poission equation by which densities of the constituent particles have been coupled.

## 4. Linear dispersion relation

In order to investigate the nonlinear behaviour of EASWs, we employ the following perturbation expansion for the field quantities $n_\alpha$, $v_\alpha$ and $\phi$ about their equilibrium values

$$\begin{pmatrix} n_\alpha \\ v_\alpha \\ \phi \end{pmatrix} = \begin{pmatrix} n_{0\alpha} \\ v_{0\alpha} \\ \phi_0 \end{pmatrix} + \varepsilon \begin{pmatrix} n_\alpha^{(1)} \\ v_\alpha^{(1)} \\ \phi^{(1)} \end{pmatrix} + \varepsilon^2 \begin{pmatrix} n_\alpha^{(2)} \\ v_\alpha^{(2)} \\ \phi^{(2)} \end{pmatrix}.$$

(14)

Substituting the above expansion (14) in eq.(6) and eqs.(11)-(13), and then taking their linear terms (liniearizing) with assumption that all the field quantities vary as $e^{i(kx-\omega t)}$, where $k$ is the wave number and $\omega$ is wave frequency, we get

$$v_\alpha^{(1)} = \frac{\omega n_\alpha^{(1)}}{k n_{0\alpha}},$$

(15)

$$v_c^{(1)} = \frac{H_{ce}^2 v_{Fce}^4}{4\omega_{Pce}^2 n_{ce0}} \frac{k^3}{\omega} n_{ce}^{(1)} - \frac{ke}{\omega m_e} \phi^{(1)},$$

(16)

$$0 = \frac{e}{m_e} \phi^{(1)} - \left( \frac{H_{he}^2 v_{Fhe}^4}{4\omega_{Phe}^2 n_{he0}} k^2 + \frac{v_{Fhe}^2}{n_{h0}} \right) n_{he}^{(1)},$$

(17)

and

$$k^2 \phi^{(1)} = -\frac{e}{\varepsilon_0} \left[ n_{ce}^{(1)} + n_{he}^{(1)} - e\alpha_\kappa \beta \left( \frac{\kappa + s - 2}{4\kappa} \right) \right].$$

(18)

Now using the above equations (15)–(18), we have following linear dispersion relation,

$$k^2 = \left( \frac{\omega_{pce}^2}{\left( \frac{\omega^2}{k^2} - \frac{H_{ce}^2 v_{Fce}^4}{4\omega_{Pce}^2 n_{ce0}} k^2 n_{ce0} \right)} - \frac{\omega_{phe}^2}{\left( \frac{H_{he}^2 v_{Fhe}^4}{4\omega_{Phe}^2 n_{he0}} k^2 n_{he0} + v_{Fhe}^2 \right)} - \frac{e^2 \alpha_\kappa \beta}{\varepsilon_0} \left( \frac{\kappa + s - 2}{4\kappa} \right) \right).$$

(19)

This dispersion relation demonstrates how different electron populations (cold, hot and kappa) contribute to the propagation of electron acoustic waves in a plasma. The first term on the right hand side is the cold electron term which dominates at low wavenumbers and low thermal speeds, providing the main inertia for the wave. The second term on right hand side is hot electron term that adds pressure effects from the hot electrons. The last term on right hand side accounts for the presence of Kappa-electrons and their collective response which shows the thermal corrections and enhances the wave's dispersion due to the presence of the Kappa Fermi distributed electrons. The term $k^2$ on the left hand side shows the total wave number of the EASW depends on contributions from cold electrons, hot electrons and Kappa-distributed electrons i.e. each component affects the wave differently.

In the numerical analysis to follow, the parameters are chosen for dense astrophysical plasmas like white dwarfs and neutron stars, having values in range of $T_{Fe} \approx 10^7 - 10^8 K$, $10^{28} \leq n_{e0} \leq 10^{32} m^{-3}$ so that $0.24 \leq H \leq 1.10$ [57,58] and the value of kappa parameter $\kappa$ for such type of dense astrophysical objects ranges from 0.1 to 0.5.

Fig. 1 shows dispersion curve for different values of quantum parameter $H = H_{ce}$ as 0.51, 0.75 and 1.1. As $H$ increases the slope of $kc/\omega_p$ vs $\omega/\omega_p$ plot increases i.e., the wavenumber increases more steeply with the normalized frequency as the quantum parameter increases. However, $H$ is directly related to quantum effects such as Fermi pressure and the Bohm potential. These effects shift the dispersion curve upward and enhance the propagation characteristics of plasma waves. This behaviour is characteristic of plasmas in highly quantum dominated regimes, where quantum pressure provides the stability and leads to steeper dispersion curves.

Fig. 2 shows dispersion curve for different values of Kappa index $\kappa$ keeping other parameters constant. As the value of kappa index $\kappa$ increases the slope of $kc/\omega_p$ vs $\omega/\omega_p$ plot increases, implying a stronger dependence of the wavenumber on the frequency at larger Kappa index. So, it is confirmed that the normalized wave vector $kc/\omega_p$ decreases with reduction in Kappa-Fermi distributed electrons. However, The Kappa parameter describes a deviation from standard Fermi form as well as presence of internal correlations. It increases the quantum domain slightly into the higher energy / temperature range by shifting Fermi energy, preferably for small $\kappa$. Hence, this parameter changes the overall energy balance and also modifies the wave dispersion characteristics. Higher $\kappa$ values shows more resemblance with standard Fermi forms i.e., higher value of $\kappa$ reduces the effective thermal speed of the particles which decreases the internal correlations and the wave-particle interactions that support wave propagation. In contrast for low-kappa index, the plasma behaves more thermally, allowing stronger collective interactions and supporting larger wavenumbers for the same frequency. Thus, we observe that as the standard Fermi distribution is approached, the dependence of the wavenumber on the frequency significantly increases and the model where the hot Kappa-Fermi distributed electrons are dominated by the standard Fermi distributed cold electrons comes into account.

Fig. 3 shows the dispersion curve for different values of density $n_{\kappa e0}$. Increasing $n_{\kappa e0}$ means more initial Kappa electrons as compared to Fermi-electrons. A greater number of $n_{\kappa e0}$ implies a increase in $kc/\omega_p$ vs $\omega/\omega_p$ slope, implying a stronger dependence of the normalized wavenumber $kc/\omega_p$ on the normalised frequency at larger Kappa electrons number density, while phase speed $\omega/k$ of the wave decreases gradually. However, electron number density strongly influences the plasma frequency. A higher kappa electron density increases the plasma's oscillatory frequency, leading to stronger collective interactions between charged particles and enhances wave propagation characteristics. This results in an increased plasma frequency and higher values for the same $\omega/\omega_p$.

## 5. Soliton solution

In order to find the solitary wave solution, we obtain the KdV equation employing the reductive perturbation technique (RPT). According to RPT, the independent variables $x$ and t are stretched as

$$\zeta = \varepsilon^{1/2}(x - \lambda t)$$

and

$$\tau = \varepsilon^{3/2} t,$$

where, $\varepsilon$ is the strength of nonlinearity and $\lambda$ is the phase velocity of the wave. The parameter, $\varepsilon$ may be interpreted as the size of the perturbation. Here, $x$ and $t$ are functions of $\zeta$ and $\tau$, so partial derivatives with respect to $x$ and $t$ can be transformed into partial derivative in terms of $\zeta$ and $\tau$,

$$\frac{\partial}{\partial x} = \varepsilon^{1/2} \frac{\partial}{\partial \zeta},$$

$$\frac{\partial}{\partial t} = -\varepsilon^{1/2} \lambda \frac{\partial}{\partial \zeta} + \varepsilon^{3/2} \frac{\partial}{\partial \tau},$$

$$\frac{\partial^2}{\partial x^2} = \varepsilon \frac{\partial^2}{\partial \zeta^2},$$

and

$$\frac{\partial^3}{\partial x^3} = \varepsilon^{3/2} \frac{\partial^3}{\partial \zeta^3}.$$

Now writing the eqs. (6) and (11) - (13) in terms of stretched co-ordinates $\zeta$ and $\tau$, we obtain

$$0 = -\varepsilon^{1/2} \lambda \frac{\partial n_\alpha}{\partial \zeta} + \varepsilon^{3/2} \frac{\partial n_\alpha}{\partial \tau} + \varepsilon^{1/2} \frac{\partial}{\partial \zeta}(n_\alpha v_\alpha), \tag{20}$$

$$-\varepsilon^{1/2} \lambda \frac{\partial v_c}{\partial \zeta} + \varepsilon^{3/2} \frac{\partial v_c}{\partial \tau} + \varepsilon^{1/2} v_c \frac{\partial}{\partial \zeta} v_c = \frac{e}{m_e} \varepsilon^{1/2} \frac{\partial \phi}{\partial \zeta} + \frac{H_{ce}^2 v_{Fce}^4}{4\omega_{Pce}^2 n_{ce0}} \varepsilon^{3/2} \frac{\partial^3 n_{ce}}{\partial \zeta^3}, \tag{21}$$

$$0 = \frac{e}{m_e} \varepsilon^{1/2} \frac{\partial \phi}{\partial \zeta} + \frac{H_{he}^2 v_{Fhe}^4}{4\omega_{Phe}^2 n_{he0}} \varepsilon^{3/2} \frac{\partial^3 n_{he}}{\partial \zeta^3} - \frac{v_{Fhe}^2}{n_{he0}^2} \varepsilon^{1/2} \left( n_{he} \frac{\partial n_{he}}{\partial \zeta} \right), \tag{22}$$

and

$$\varepsilon \frac{\partial^2 \phi}{\partial \zeta^2} = \frac{e}{\varepsilon_0} \left[ n_{ce} + n_{he} - Z_i n_{i0} + \alpha_\kappa \left\{ 1 - \left( \frac{\kappa + s - 2}{4\kappa} \right) \beta \right\} \right]. \tag{23}$$

Applying the perturbative expansion eq. (14) in the above transformed set of equations and taking the lowest order terms of $\varepsilon$, we get the following set of equations

$$n_{0\alpha} \frac{\partial v_\alpha^{(1)}}{\partial \zeta} - \lambda \frac{\partial n_\alpha^{(1)}}{\partial \zeta} = 0, \tag{24}$$

$$\frac{e}{m_e} \frac{\partial \phi^{(1)}}{\partial \zeta} + \lambda \frac{\partial v_c^{(1)}}{\partial \zeta} = 0, \tag{25}$$

$$\frac{e}{m_e} \frac{\partial \phi^{(1)}}{\partial \zeta} - \frac{v_{Fhe}^2}{n_{0h}} \frac{\partial n_{he}^{(1)}}{\partial \zeta} = 0, \tag{26}$$

and

$$\frac{e}{\varepsilon_0} \left[ n_{ce}^{(1)} + n_{he}^{(1)} - e\alpha_\kappa \beta \left( \frac{\kappa + s - 2}{4\kappa} \right) \phi^{(1)} \right] = 0. \tag{27}$$

Solving these equations we get following perturbed quantities

$$n_c^{(1)} = \frac{n_{ce0}}{\lambda} v_c^{(1)},$$

$$v_c^{(1)} = -\frac{e}{m_e \lambda} \phi^{(1)},$$

$$n_c^{(1)} = -\frac{e n_{ce0}}{m_e \lambda^2} \phi^{(1)}$$

and

$$n_h^{(1)} = \frac{e n_{he0}}{m_e v_{fe}^2} \phi^{(1)}.$$

We also get the phase velocity as,

$$\lambda^2 = \frac{e n_{ce0}}{m_e \left( e\alpha_\kappa \beta \left( \frac{\kappa + s - 2}{4\kappa} \right) - \frac{e n_{he0}}{m_e v_{Fhe}^2} \right)}. \tag{28}$$

Now again, substituting the expansion (14) in the set of equations (24)–(27) and taking the next higher order terms of $\varepsilon$, we obtain

$$-\lambda \frac{\partial n_\alpha^{(2)}}{\partial \zeta} + \frac{\partial n_\alpha^{(1)}}{\partial \tau} + n_{o\alpha} \frac{\partial v_\alpha^{(2)}}{\partial \zeta} \frac{\partial}{\partial \zeta} \left( n_\alpha^{(1)} v_\alpha^{(1)} \right) = 0, \tag{29}$$

$$-\lambda \frac{\partial v_c^{(2)}}{\partial \zeta} + \frac{\partial v_c^{(1)}}{\partial \tau} + v_c^{(1)} \frac{\partial v_c^{(1)}}{\partial \zeta} = \frac{e}{m_e} \frac{\partial \phi^{(2)}}{\partial \zeta} + \frac{H_{ce}^2 v_{Fce}^4}{4 \omega_{Pce}^2 n_{ce0}} \frac{\partial^3 n_{ce}^{(1)}}{\partial \zeta^3}, \tag{30}$$

$$\frac{e}{m_e} \frac{\partial \phi^{(2)}}{\partial \zeta} - \frac{v_{Fhe}^2}{n_{0h}^2} \frac{\partial n_{he}^{(2)}}{\partial \zeta} - \frac{v_{Fhe}^2}{n_{0h}^2} \left( n_{he}^{(1)} \frac{\partial n_{he}^{(1)}}{\partial \zeta} \right) + \frac{H_{he}^2 v_{Fhe}^4}{4 \omega_{Phe}^2 n_{eho}} \frac{\partial^3 n_{he}^{(1)}}{\partial \zeta^3} = 0, \tag{31}$$

and

$$\frac{\partial^2 \phi^{(1)}}{\partial \zeta^2} = \frac{e}{\varepsilon_0}\left[ n_c^{(2)} + n_{he}^{(2)} - e\alpha_\kappa \left(\frac{\kappa+s-2}{4\kappa}\right)\beta\phi^{(2)} \right].$$

(32)

Solving these equations we get KdV equation as,

$$\frac{\partial \phi^{(1)}}{\partial \tau} + C_1 \phi^{(1)} \frac{\partial \phi^{(1)}}{\partial \zeta} + C_2 \frac{\partial^3 \phi^{(1)}}{\partial \zeta^3} = 0$$

(33)

where, $C_1 = \frac{e\lambda}{2m_e}\left[\left(\frac{\omega_{phe}\lambda}{\omega_{pce}v_{Fhe}^2}\right)^2 - \left(\frac{\sqrt{3}}{\lambda}\right)^2\right]$, and $C_2 = \frac{\lambda}{2}\left[\left(\frac{\lambda}{\omega_{pce}}\right)^2 - \frac{n_{ce0}Q_{ce}}{\lambda^2} - \frac{H_{he}^2 v_{Fhe}^4}{4\omega_{Phe}^2}\left(\frac{\omega_{phe}\lambda}{\omega_{pce}v_{Fhe}^2}\right)^2\right].$

The second term in the eq. (33) is the nonlinear term and the last term is the dispersive terms. $C_1$ and $C_2$ are nonlinear and dispersive coefficients respectively. Nonlinearity can transfer energy into the wave, leading to wave breaking. However, the combined presence of nonlinearity and dispersion allow for the formation of a stable wave profile. The steady-state solution of this KdV equation is obtained by transforming the independent variables $\zeta$ and $\tau$ to $\eta = \zeta - U\tau$ where, $U$ is a normalised constant speed of electron acoustic wave frame. Applying the boundary condition that as $\eta \to \pm\infty$, $\phi \to 0$, $\partial\phi/\partial\eta \to 0$ and $\partial^2\phi/\partial\eta^2 \to 0$, the possible stationary solution of the KdV eq. (33) is obtained as

$$\phi = \phi_m \operatorname{sech}^2\left(\frac{\eta}{\Delta}\right),$$

(34)

where, the amplitude $\phi_m$ and width $\Delta$ of solitary waves are

$$\phi_m = \frac{3U}{C_1}, \text{ and } \Delta = \sqrt{\frac{4C_2}{U}}.$$

The formation of the solitary wave structure arises from the balance between the dispersive and nonlinear terms. The coefficients $C_1$ and $C_2$ play a crucial role in determining the solitary wave structure, with their nature and magnitude primarily determining the characteristics of the resulting soliton. Therefore, it is essential to examine how the nonlinear and dispersive coefficients depend on various physical plasma parameters. The quantum effect influences only the dispersive coefficients. Both these coefficients depend on $\Delta$, the equilibrium cold-to-hot electron concentration ratio. Specifically, while the nonlinear coefficient $C_2$ depends both on $\Delta$ and quantum parameter $H$, the dispersive coefficient $C_1$ depends only on $\Delta$.

Fig. 4 shows the variation in the electron-acoustic solitary wave solution of the KdV equation for different values of Mach number $U$. It is observed that as the normalized phase speed $U$ increases, the amplitude $\phi_m$ of the solitary wave increases, but the width $\Delta$ decreases, thus causes the amplitude to grow i.e., the solitary wave becomes spiky. However, the width of the soliton inversely depends on $U$. As $U$ increases, the wave becomes more localized due to stronger nonlinearity, leading to narrower solitons. Higher-speed electron acoustic waves carry more energy, compressing the wave spatially while amplifying its peak.

Fig. 5 illustrates the behaviour of electron-acoustic solitary wave solutions of the KdV equation as a function of the kappa index, while keeping other parameters constant. The results show that an increase in the kappa index leads to an increase in the amplitude of the solitary wave and a reduction in its width, making the wave more sharply peaked. This spiky nature emerges because Kappa-distributed hot electrons, as they approach the standard Fermi distribution (higher kappa values), enhance the nonlinearity of the system while diminishing dispersion effects. Consequently, the soliton becomes more localized. In contrast, lower kappa indices result in weaker solitary wave structures with broader profiles, as the system exhibits stronger dispersion and less pronounced nonlinear effects.

Fig. 6 shows the variation in the electron-acoustic solitary wave solution of the KdV equation for different values of density $n_{\kappa e0}$ with fixed values of other parameters. It is seen that as the equilibrium density of kappa electrons increases, the amplitude and width of the solitary wave decrease, resulting in narrower and taller profiles due to stronger nonlinearity and weaker dispersive property.

## 6. Summary and discussion

Linear and non linear Electron acoustic solitary waves (EAWs) have been investigated in the presence of Kappa-Fermi distributed electrons for astrophysical quantum plasma comprising cold and hot electron fluid with stationary ions. A dispersion relation for linear EAW have been derived using QHD model incorporating the quantum effects of electron's Fermi pressure, the quantum Bohm potential by using non-dimensional quantum diffraction parameter. An analytical solution for the EASW has been derived employing the standard RPT, and deriving the KdV equation for the EASWs. The effects of the Kappa Fermi distributed hot electron number densities at equilibrium ($n_{\kappa e0}$), kappa index ($\kappa$) and quantum parameter ($H$) on dispersion properties of EAW while speed of the traveling wave ($U$), $\kappa$, and $n_{\kappa e0}$ have been investigated on the analytical solution of the EA solitary waves.

The dispersion characteristics reveal that increasing $H$ steepens the slope of the wave number ($k$) versus frequency ($\omega$) due to quantum effects like Fermi pressure and Bohm potential, which enhance the plasma's restoring force and wave propagation. Similarly, higher κ values, representing a system closer to the standard Fermi distribution, reduce wave-particle interactions and thermal speed, leading to steeper dispersion curves. Electron density ($n_{\kappa e0}$) increases the plasma frequency and collective interactions, resulting in stronger wave coherence and steeper dispersion behaviour. These findings emphasize the significant roles of quantum and statistical effects in shaping wave dynamics in quantum plasma. The solitary wave analysis demonstrates that higher Mach numbers ($U$) increase the amplitude and reduce the width of solitons, indicating stronger nonlinearity and energy localization. Similarly, increasing κ enhances localization, producing taller and narrower solitons as the plasma transitions toward a Fermi-like distribution. For higher $n_{\kappa e0}$, solitons become narrower and taller due to stronger nonlinear coupling and increased collective interactions. These results highlight the intricate interplay between dispersion and nonlinearity, showing how various plasma parameters govern wave propagation and soliton characteristics. Our analysis reveals the influence of Kappa-Fermi distributed electron populations and the effects of electron's Fermi pressure and the quantum Bohm potential on the wave dynamics, providing insight into the behaviour of EA solitons in astrophysical bodies like white dwarf and neutron stars. This study provides a deeper understanding of the quantum effects, statistical properties, and nonlinear wave behaviour in dense astrophysical plasma as well as provides a basis for exploring practical applications in plasma based technologies.

**Declaration of competing interest**

The authors declare that they have no known competing financial interests or personal relationships that could have appeared to influence the work reported in this paper.

**Data availability**

No data was used for the research described in the article.

**Acknowledgement**

The authors thank SERB- DST, Govt. of India for financial support under MATRICS scheme (grant no. : MTR/2021/000471).

# Figure captions

Fig.1     Variation of $kc/\omega_p$ with $\omega/\omega_p$, for different values of quantum parameter $H$ with $n_{\kappa e0} = 10^{29} m^{-3}$ and $\kappa = 0.5$

Fig.2     Variation of $kc/\omega_p$ with $\omega/\omega_p$, for different values of Kappa index $\kappa$ with $H = 0.747$ and $n_{\kappa e0} = 10^{29} m^{-3}$

Fig.3     Variation of $kc/\omega_p$ with $\omega/\omega_p$, for different values of equilibrium kappa number density $n_{\kappa e0}$ with $H = 0.747$ and $\kappa = 0.5$

Fig.4     Variation of the solitary wave solution $\phi$ with $\zeta$, for the different values of $U$ with $n_{\kappa e0} = 10^{29} m^{-3}$ and $\kappa = 0.5$

Fig.5     Variation of the solitary wave solution $\phi$ with $\zeta$, for the different values of Kappa index $\kappa$ with $U = 0.1$ and $n_{\kappa e0} = 10^{29} m^{-3}$

Fig.6     Variation of the solitary wave solution $\phi$ with $\zeta$, for the different values of number density $n_{\kappa e0}$ with $U = 0.1$ and $\kappa = 0.5$

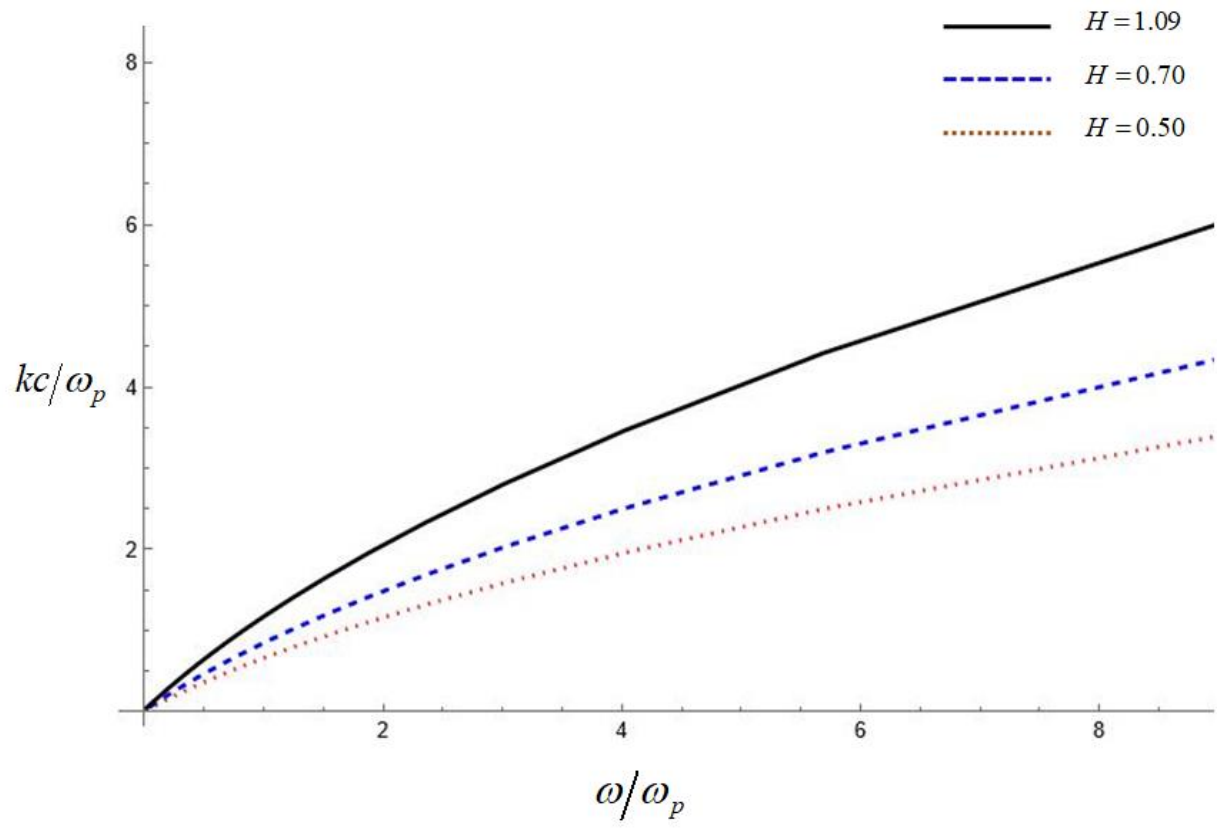

Fig.1

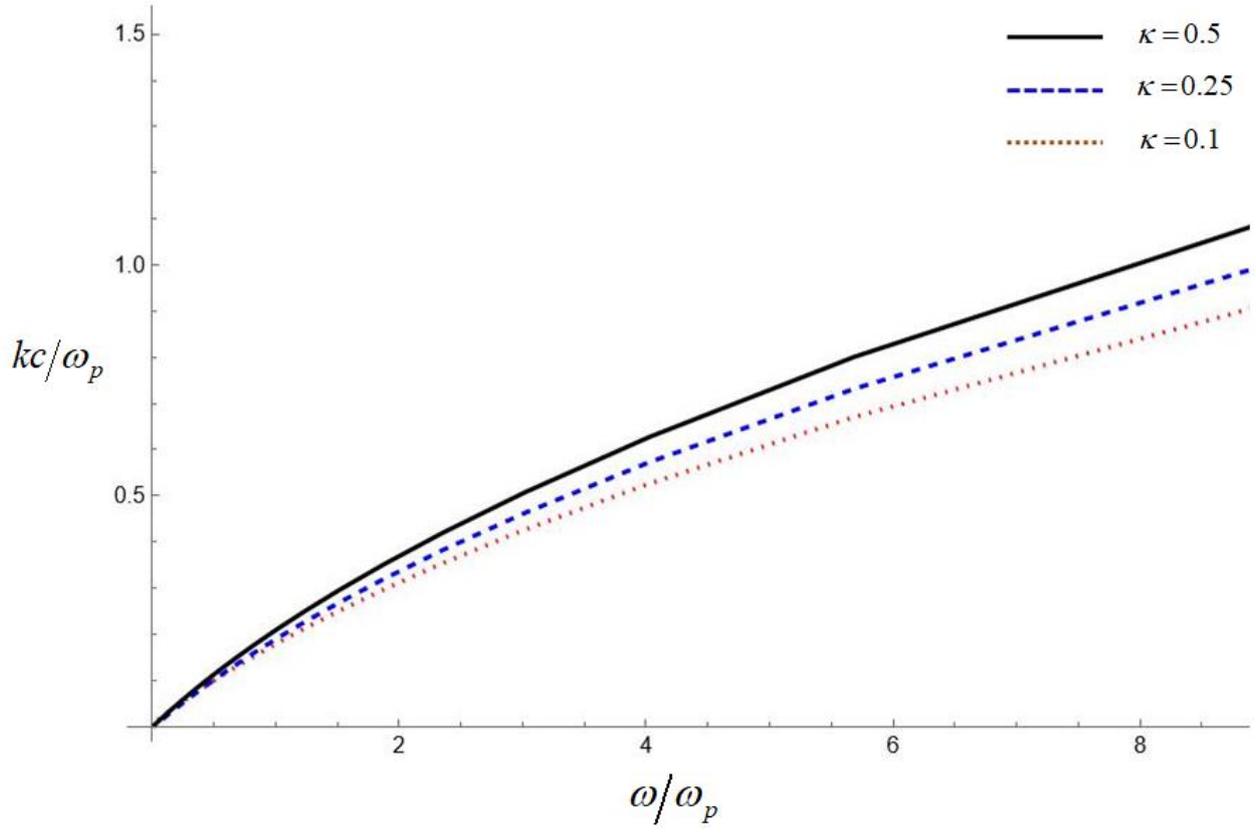

Fig.2

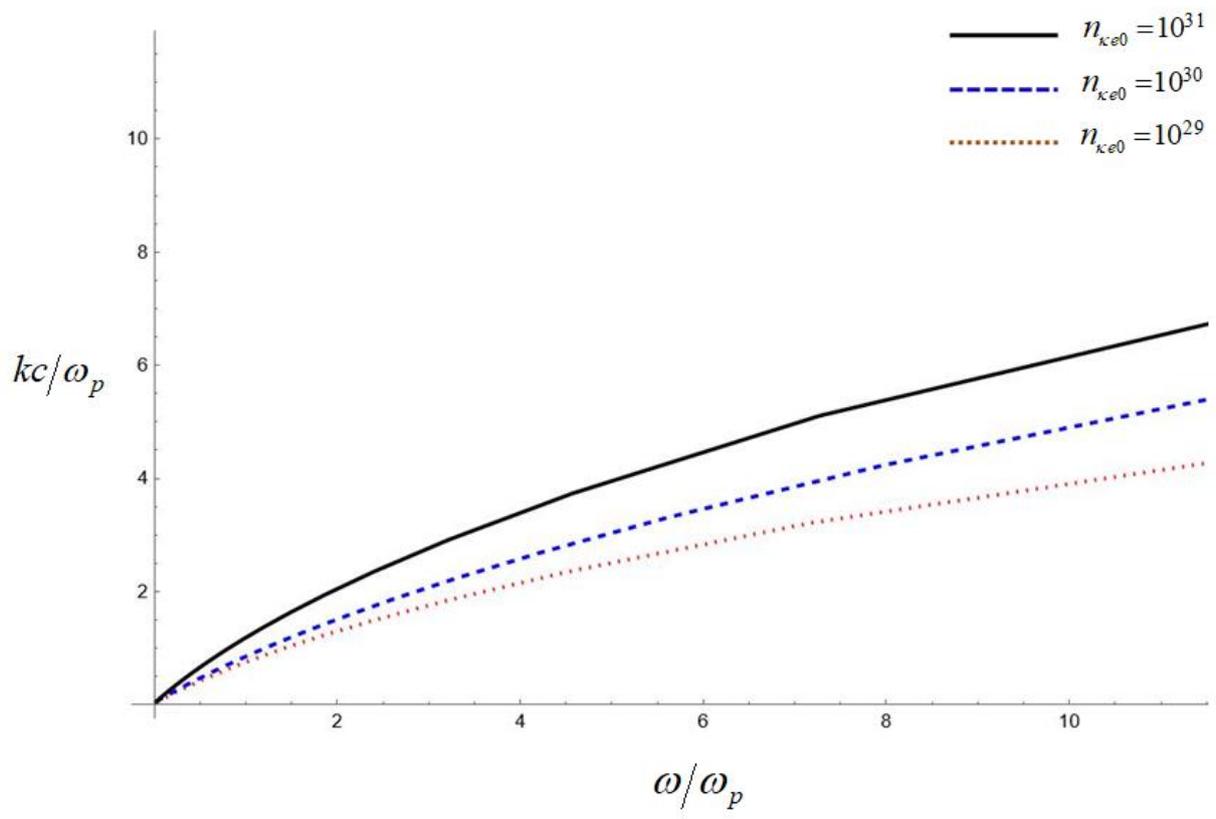

Fig.3

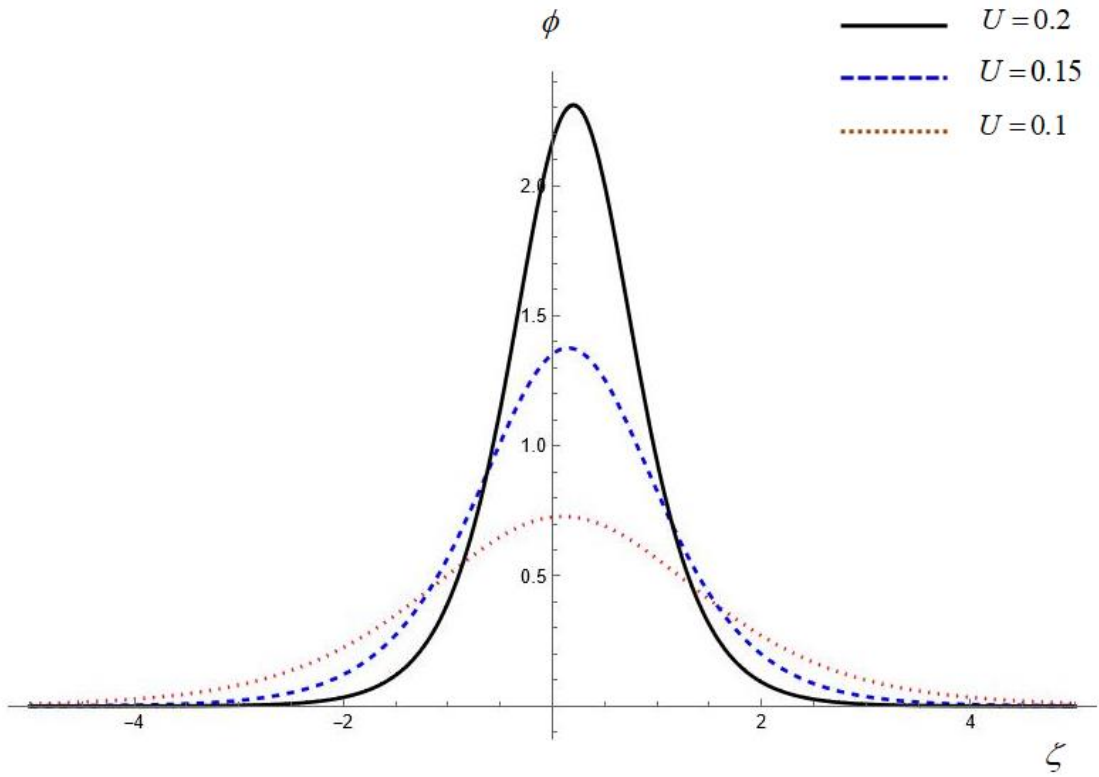

Fig.4

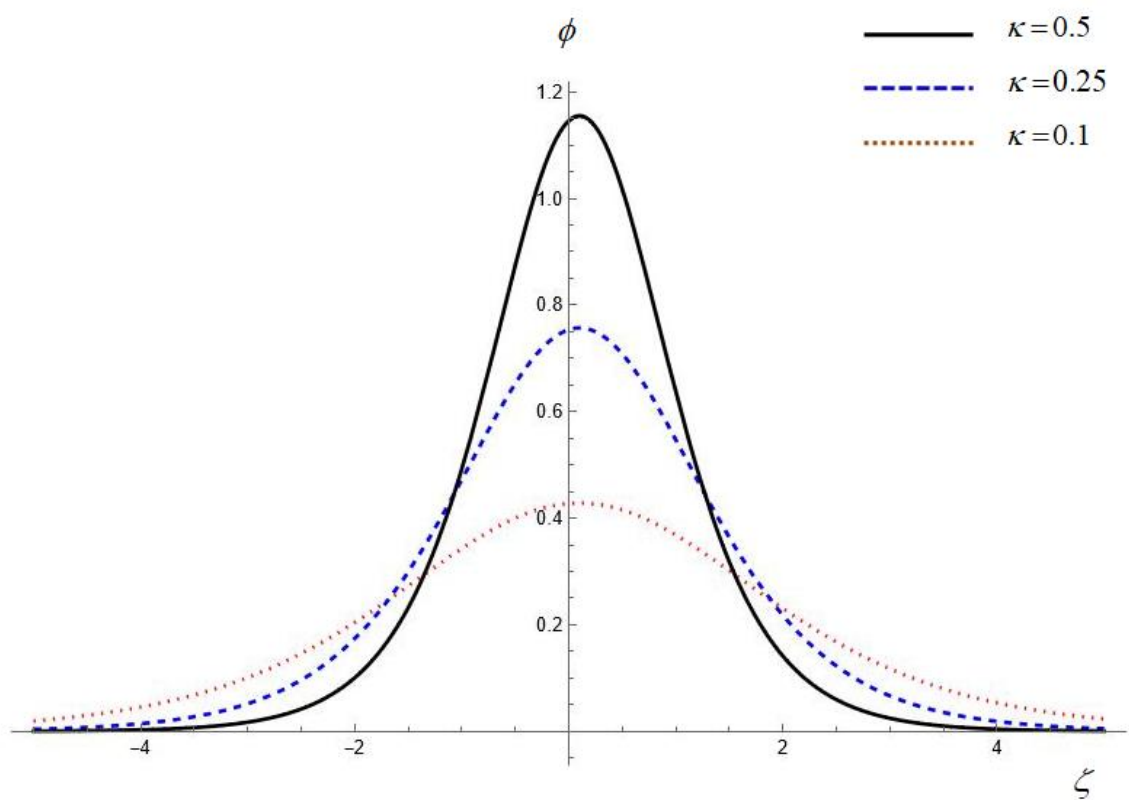

Fig.5

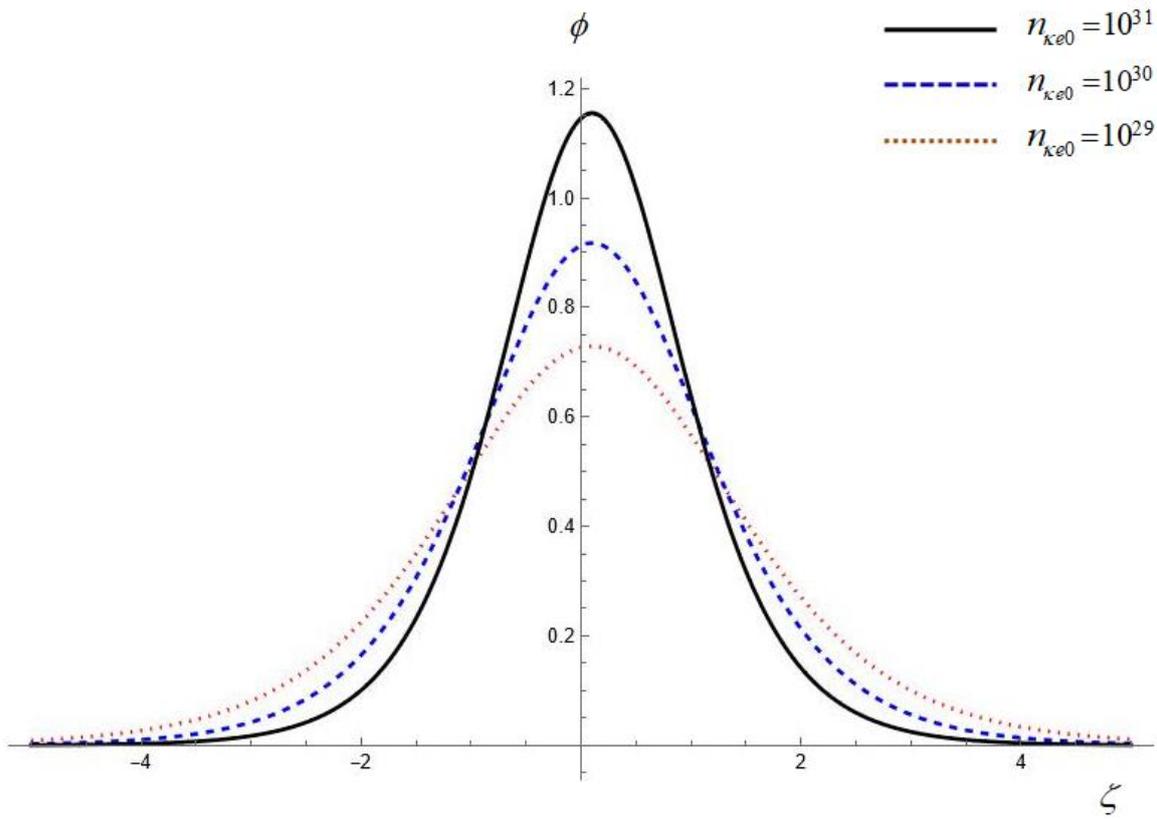

Fig.6